\documentclass[conference]{IEEEtran}
\usepackage{cite}
\usepackage{amsmath,amssymb,amsfonts}
\usepackage{algorithmic}
\usepackage{graphicx}
\usepackage{textcomp}
\usepackage{xcolor}
\usepackage{fancyhdr}
\usepackage{comment}
\usepackage{todonotes}
\usepackage{lipsum}
\usepackage{makecell}
\usepackage[hyphens]{url}

\def\BibTeX{{\rm B\kern-.05em{\sc i\kern-.025em b}\kern-.08em
    T\kern-.1667em\lower.7ex\hbox{E}\kern-.125emX}}

\title{Workload Similarity Analysis using Machine Learning Techniques} 
\author{
\IEEEauthorblockN{Ashish Ledalla }
\IEEEauthorblockA{IIT Jodhpur, India\\
Email: ledalla.1@iitj.ac.in}
\and
\IEEEauthorblockN{Vineet Singh}
\IEEEauthorblockA{Intel Corporation, Hillsboro, USA\\
Email: vineet.singh@intel.com }
\and
\IEEEauthorblockN{Deepak Mishra}
\IEEEauthorblockA{IIT Jodhpur, India\\
Email: dmishra@iitj.ac.in}
}
\begin{document}
\maketitle
\pagestyle{plain}


\begin{abstract}
Finding the similarity between two workload behaviours is helpful in 1. creating proxy workloads 2. characterising an unknown workload's behavior by matching its behavior against known workloads. In this article, we propose a method to measure the similarity between two workloads using machine learning based analysis of the performance telemetry data collected for the execution runs of the two workloads. We also demonstrate the accuracy of the technique by measuring the similarity between a variety of know benchmark workloads.
\end{abstract}

\section{Introduction}
Optimizing the performance of applications along with the efficient utilization of underlying computing resources is an important problem that require high level of expertise to resolve. Workloads can vary from HPC workloads running on supercomputers to SasS(Software as a Service) workloads running in cloud. This vast variety of workloads and underlying computing resources like examples of computing resources brings different sets of optimization challenges, making workload optimization a very time and skill intensive process. Being able to automate any part of performance and resource optimization makes a signification impact on cost and turn around time of optimization.

One of the crucial sub-problems for performance and resource utilization optimization of real-world workloads is generating proxy-workloads ~\cite{panda}. Owing to the availability and reproducibility of real-world workloads,   proxy-workloads act as a replacement to perform same analysis and optimization. Proxy-workloads have similar computing resource requirements to that of the original workload and optimizations done of proxy-workloads have a high applicability to the corresponding real-world workload.

Another important step in optimizing a workload is characterizing the computing resource utilization behavior of the workload~\cite{spec2017}. Once the workload behaviour is identified to be similar to well-know resource utilization patterns, the standard methods for optimization  applicable to the known behavior can be applied to optimize the workload performance. For example, once the user has identified that the workload is memory size intensive, a machine with additional memory space can be used for running the workload to get the best performance. Characterizing an unknown workload is time consuming task if done manually.

Both the problems of generating proxy-workloads and characterizing an unknown workload can be modelled as similarity analysis of two given workloads. Similarity analysis is defined as matching the behaviour of two given workloads and calculating their similarity in terms of how they are using the computing resource and the way both workloads stress the computing resources.

In this article, we propose a workload similarity analysis technique which matches the telemetry data collected from running both the input workloads to calculate a similarly index. The telemetry data is matched based on multi-variate time series (MTS) matching proposed in \cite{yang2004pca}.
The telemetry data collected over the execution of the workload is referred as the signature of the workload. We have used a data collected using 1. SYSSTAT (SAR)\cite{SAR} and 2. Linux perf \cite{perf} as the signature of the workload.

Comparing the signatures of any two workloads is not trivial as they are of varying lengths in terms of the number of rows. For this reason, we make use of the methodology given in \cite{yang2004pca} where MTS samples with varying lengths can be compared based on their proposed similarity measure called Eros \cite{yang2004pca}.

Once we collect the signature of the input workload, we use the Eros similarity as a measure to compare it with another workload's signature. For instance, the given workload can be identified as how close or far away it is from a known workload based on the magnitude of the similarity value in terms of its behaviour and characteristics. Based on the similarity value, the unknown workload can be analyzed further based on the properties of the known signature.

Rest of the paper is organised as following. Section~\ref{sec:bck} describes the background of workload similarity analysis and other approaches. Section~\ref{sec:met} describes our similarity analysis methodology in details. Section~\ref{sec:res} provides and describes the results of our experiments. 

\section{Background and Related work}
\label{sec:bck}
In computing world, workload is an applications or program that when run on a computer system uses the system computing resources. These computing resources include but are not limited to CPU, memory, storage and network resources. The performance of a workload is the amount of useful work done during the execution of the workload in a given time.

The problem boils down to time-series similarity measures where there are several techniques based on Principle Component Analysis, Dynamic Time Warping, Mahalanobis distance etc. The possibility of having variable length time-series with complex inter-dependencies among different variables poses the challenge for selecting a good similarity measure. There are also recent advances in deep learning that focus on deep metric learning for comparing and measuring the similarity between time-series data. Recent works on deep metric learning like DECADE\cite{che2017decade} which is based on an innovative alignment technique called expected alignment gives a data dependent local representation using end to end gradient learning. Another recent work \cite{franceschi2019unsupervised} inspired from word2vec is based on unsupervised deep learning utilizing a novel form of triplet loss in formulating general-purpose representations that are scalable irrespective of the length of the MTS. Recent advances in self-supervised learning for representation learning are pushing the research in MTS similarity analysis, for instance, in \cite{eldele2021time} the authors proposed a novel self-supervised technique (TS-TCC) that is based on weak and strong augmentations of the given time-series data and learns the representations from the unlabelled data via a cross-view prediction task. These are some of the recent methods in the context of deep-metric learning that seem promising for our work.

The challenge with these above-mentioned techniques is that they require huge amounts of data for training the deep learning based models. Another important aspect is the parameters of the deep learning model need to be tweaked (model has to be retrained with the new examples) every time a new telemetry MTS item that the model has not seen before is added to the database. In our case, the data currently available for the analysis is not sufficient to train a deep learning model and thus we preferred a more simpler similarity measure. PCA-based Eros similarity measure is a linear model and less time taking in terms of obtaining the model parameters. Though this is a relatively old technique, it is computationally inexpensive and acts as a good prototype technique for MTS similarity measure and thus we chose to use this for our preliminary experiments on our limited database of MTS data.

\section{Methodology}
\label{sec:met}
In this section we introduce our methodology of finding similarity between two workloads. Our approach is focused on the similarity of the workloads based on their usage of computing resources on the machine. The first part of our approach is to define and collect the signature of the workload that represents workload's usage of the machine computing resources. In the second part we measure the similarity between the signatures of two workloads using Eros similarity measure.
    The rest of the section describes the two parts of our approach in details.

\begin{scriptsize}
    \begin{table}[h!]
    \centering
    \caption{The signature metrics and their definitions collected using SYSSTAT (SAR) ~\cite{SAR}}
    \label{tab:os_metrics}
    \begin{tabular}{|p{2cm}|p{5cm}|}
    \hline
    \textbf{Event/Metric Name} & \textbf{Description} \\
    \hline
    tps (IOPS) & Transfers per second issued to physical devices. \\
    \hline
    Rtps (read IOPS) & Read requests per second issued to physical devices. \\
    \hline
    wtps (write IOPS) & Write requests per second issued to physical devices. \\
    \hline
    bread/s & Blocks per second read\\
    \hline
    bwrtn/s & Blocks per second written.\\
    \hline
    pgpgin/s & Kilobytes per second paged in \\
    \hline
    pgpgout/s & Kilobytes per second paged out\\
    \hline
    fault/s & Page faults (major + minor) per second\\
    \hline
    majflt/s & Major page faults per second\\
    \hline
    rkB/s & Kilobytes per second read from the device \\
    \hline
    wkB/s & Kilobytes per second written to the device \\
    \hline
    areq-sz & Average size (in kilobytes) of the I/O requests\\
    \hline
    aqu-sz & Average queue length of the I/O requests\\
    \hline
    await & Average time (in milliseconds) for I/O requests issued to the device to be served\\
    \hline
    kbmemfree & Free memory available in kilobytes.\\
    \hline
    kbmemused & Used memory in kilobytes (installed memory - kbmemfree - kbbuffers - kbcached - kbslab).\\
    \hline
    kbbuffers & Memory used as buffers by the kernel in kilobytes.\\
    \hline
    kbslab & Memory in kilobytes used by the kernel to cache data structures for its own use.\\
    \hline
    \%user & Percentage of CPU utilization that occurred while executing at the user level (application)\\
    \hline
    \%system & Percentage of CPU utilization that occurred while executing at the system level (kernel)\\
    \hline
    \%iowait & Percentage of time that CPUs were idle during which the system had an outstanding disk I/O request.\\
    \hline
    \%idle & Percentage of time that CPUs were idle and the system did not have an outstanding disk I/O request.\\
    \hline
    pswpin/s  & Swap in pages per second.\\
    \hline
    pswpout/s & Swap out per second. \\
    \hline
    IFACE & Name of the network interface for which statistics are reported.\\
    \hline
    rxkB/s & Kilobytes received per second\\
    \hline
    txkB/s  & Kilobytes transmitted per second.\\
    \hline
    \%ifutil & Utilization percentage of the network interface\\
    \hline
    \end{tabular}

    \end{table}
    \end{scriptsize}

\begin{scriptsize}
    \begin{table}[h!]
    \centering
    \caption{The telemetry metrics and their definitions collected using Linux Perf~\cite{perf} }
    \label{tab:hw_metrics}
    \begin{tabular}{|p{2cm}|p{5cm}|}
    \hline
    \textbf{Metric Name} & \textbf{Description} \\
    \hline
    branches & All branch instructions retired\\
    \hline
    L1 data cache-references & Number of L1 data cache references.\\
    \hline
    L1 data cache-misses & Number of L1 data cache-misses\\
    \hline
    cycles & Core cycles when the thread is not in halt state.\\
    \hline
    instructions & The number of instructions retired from execution.\\
    \hline
    ref-cycles & Reference cycles when the core is not in halt state.\\
    \hline
    \end{tabular}

    \end{table}
\end{scriptsize}

\subsection{Workload signature definition and collection}
We have defined the signature of a workload to be a the multivariate time series of performance metrics. Workload signature represents the run of the workload on a given machine. The metrics used measure the uses of different components of the machine by workload. Table ~\ref{tab:os_metrics} gives the list of metrics that are reported by the OS measuring cpu usage, memory usage, I/O, network, swapping, page faults behavior of the workload. Table ~\ref{tab:hw_metrics} lists the hardware events that we use as workload signature. Hardware events show the cycles, instructions, branches and cache behavior of the workload.
We use SYSSTAT (SAR) ~\cite{SAR} linux utility to collect OS based performance metrics. We collect the hardware events using Linux Perf ~\cite{perf}. Both the collectors are open-source and available with standard linux distributions.

\begin{enumerate}
    \item \textbf{SYSSTAT (SAR):~\cite{SAR}}
    It is a sysstat utility tool used to monitor Linux system/subsystem performance analytics. We can analyze/collect performance data in real-time and also store the data. Basically, it’s a Linux performance statistics analyzer.
        \item \textbf{Linux Perf (in linux-tools-generic):\cite{perf}}
    Perf is a simple but powerful performance monitoring tool for Linux-based operating systems. It is used to trace or count both hardware and software events. It provides a number of subcommands and is capable of statistical profiling of the entire system.
    \end{enumerate}
 Using the tools mentioned above, we run the workload and collect the performance telemetry data in form of a multi-variate time series. This time-series data is the workload signature that we use for similarity analysis.

We used the following set of benchmarks for our experiments:
\begin{enumerate}
\item \textbf{Intel Memory Latency Checker:~\cite{MLC}} It is a small application that measures the latencies of elements/devices communicating with CPU and RAM also between CPU and RAM.\\
\item \textbf{STREAM:~\cite{stream}} The STREAM benchmark is a simple synthetic benchmark program that measures the memory bandwidth (in MB/s) and computation rate for simple vector kernels over time.\\
\item \textbf{NAS Parallel Benchmarks:~\cite{NAS}} The NAS Parallel Benchmarks (NPB) are a group of programs meant to aid in the evaluation of parallel supercomputer performance. The benchmarks are made up of five kernels and three pseudo-applications drawn from computational fluid dynamics (CFD) applications.
\end{enumerate}

\begin{scriptsize}
    \begin{table}[h!]
    \centering
    \caption{Workloads used and their definitions}
    \label{tab:workload_def}
    \begin{tabular}{|p{1.7cm}|p{5cm}|}
    \hline
    \textbf{Workload} & \textbf{Description} \\
    \hline
    BT-B (NPB) & Block Tri-diagonal solver (Class B)\\
    \hline
    BT-C (NPB) & Block Tri-diagonal solver (Class C)\\
    \hline
    CG-C (NPB) & Conjugate Gradient, irregular memory access and communication (Class C)\\
    \hline
    FT-C (NPB) & Discrete 3D fast Fourier Transform, all-to-all communication (Class C).\\
    \hline
    LU-B (NPB) & Lower-Upper Gauss-Seidel solver (Class B)\\
    \hline
    LU-C (NPB) & Lower-Upper Gauss-Seidel solver (Class C)\\
    \hline
    SP-B (NPB) & Scalar Penta-diagonal solver (Class B)\\
    \hline
    SP-C (NPB) & Scalar Penta-diagonal solver (Class C)\\
    \hline
    MLC & Intel Memory Latency Checker\\
    \hline
    STREAM & Sustainable Memory Bandwidth in High Performance Computers\\
    \hline
    \end{tabular}

    \end{table}
\end{scriptsize}

\subsection{Workload Similarity Analysis}

We use the data collected using the performance utility tools over the selected benchmark workloads in real-time to identify the similarity among the different workloads. Different workloads run for different periods of time and hence the time series data collected for each benchmark is of different length. Each workload's telemetry is a multivariate time series data (MTS) with varying lengths, 
It is, therefore, not straightforward to compare the data in its original form for different workloads. In order to compare the telemetry of different workloads, we consider Principal Component Analysis (PCA) based similarity measures. There are several modified algorithms for obtaining similarity based on PCA for MTS data with varying lengths. The similarity metric are modified into a distance metric for further applications.

For our analysis, we used a modified similarity metric based on PCA called Eros~\cite{yang2004pca}. Eros is a similarity metric based on the eigenvector matrices and eigenvalues for given telemetry data. Eros provides us with a distance metric that can give the distance between two eigenvector matrices. Here it is essential to note that all the eigenvector matrices must of the same dimensions, this can only be done when the data collected across all the workloads is having the same features/columns. Once we have the distance formulation, we use a modified leave-one-out KNN \cite{yang2004pca} to evaluate the telemetry of the workloads in the eigenvector matrix space. Eros is defined as follows,

\begin{equation}
\begin{aligned}
    Eros(A,B,w) &= \sum_{i = 1}^n w_i |<a_i, b_i>| \\
    &= \sum_{i = 1}^n w_i |cos\theta_i|
\end{aligned}
\end{equation}

In the above equation, we have A and B matrices which are $m_B \times n$ and $m_A \times n$ sized MTS samples. Suppose $V_A$ and $V_B$ are the right eigen vector matrices obtained by applying SVD on the covariance matrices $M_A$ and $M_B$ respectively, then we can represent $V_A = [a_1, a_2,...,a_n]$ and $V_B = [b_1,b_2,...,b_n]$ such that $a_i$ and $b_i$ are column orthonormal vectors of size $n$. Then we have $<a_i, b_i>$ as the inner product of $a_i$ and $b_i$ vectors. The weights $w_i$ are calculated using the whole database of given MTS samples as follows,
\begin{equation}
\begin{aligned}
    w_i \leftarrow f(s_{*i}),\:i:1\:to\:n &&\\
    w_i \leftarrow w_i/\sum_{j=1}^n w_i,\:i:1\:to\:n &&
\end{aligned}
\end{equation}
Say we have $N$ MTS samples in the database with $n$ variables for each, then for each MTS item, we obtain $n$ eigenvalues, then a matrix $S$ of size $n \times N$ can be obtained such that each column contains the eigenvalues of the $jth$ MTS item for $j:1\:to\:N$. The function $f(.)$ is an aggregating function like $mean(.)$ or $max(.)$ or $min(.)$ and $s_{*i}$ means $ith$ row in the $S$ matrix.

\begin{scriptsize}
\begin{table*}[h!]
    \centering
    \caption{Sample of data collected using SYSSTAT (SAR)~\cite{SAR}}
    \label{tab:os_data}
    \begin{tabular}{|r|r|r|r|r|r|r|r|r|r|r|r|r|r|r|r|r|r|}
\hline
 \textbf{timestamp} & \textbf{\%user} & \textbf{\%system} & \textbf{\%iowait} & \textbf{\%idle} & \textbf{flt/s} & \textbf{majflt/s} & \textbf{tps} & \textbf{rtps} & \textbf{wtps} & \textbf{kbmemfree} & \textbf{kbavail}	& \textbf{kbmemused} \\\hline
1632834009 &	77.99 &	13.68 &	0	& 8.33	& 409623 &	0	& 0	& 0	& 0	& 3013448 &	4885884	& 2592616 \\\hline
1632834010	& 99.88 &	0.12	& 0 &	0	& 2	& 0	& 2	& 0	& 2	& 3013252 &	4885744	& 2592784 \\\hline
1632834011 &	99.75 &	0.12 &	0	& 0 &	4 &	0 &	0	& 0	& 0	& 3013448 &	4886000 &	2592780 \\\hline
1632834012 &	99.75 &	0.25 &	0	& 0	& 0	& 0	& 0 &	0	& 0	& 3013472	& 4886024 &	2592752 \\\hline
1632834013 &	83.42	& 11.47 &	0	& 5.11 &	401580	& 0	& 0	& 0	& 0	& 3039492 &	4912104 &	2566540 \\\hline

\hline
    \end{tabular}
\end{table*}
\end{scriptsize}

\begin{scriptsize}
\begin{table*}[h!]
    \centering
    \caption{Sample of data collected using the Linux Perf~\cite{perf}}
    \label{tab:hw_data}
    \begin{tabular}{|r|r|r|r|r|r|r|r|r|}
\hline
 \textbf{timestamp} & \textbf{branches} & \textbf{branch-misses} & \textbf{bus-cycles} & \textbf{cache-misses} &  \textbf{cache-references} & \textbf{cycles} & \textbf{instructions} & \textbf{ref-cycles} \\
\hline
  1.001065 &  803845239.0 &          28057 &    21820269 &         86621 &          221675.0 & 2.954453e+09 &  4.725771e+09 & 1.635233e+09 \\
 \hline
  2.002247 &  891990979.0 &          21945 &    23759259 &          5797 &           94197.0 & 3.279459e+09 &  5.246710e+09 & 1.780986e+09 \\
  \hline
  3.003394 &  885859085.0 &          10583 &    23766099 &           507 &           10739.0 & 3.282927e+09 &  5.243537e+09 & 1.780905e+09 \\
  \hline
  4.003674 &  865312363.0 &          22218 &    23767142 &       1906472 &         2528759.0 & 3.250490e+09 &  5.069008e+09 & 1.781131e+09 \\
  \hline
  5.004925 &   30149894.0 &          18548 &    23894168 &       8848774 &         9200663.0 & 1.368922e+09 &  1.307644e+08 & 1.792173e+09 \\
\hline
    \end{tabular}

\end{table*}
\end{scriptsize}

Finally we have $Eros\:distance$ formulated using the $Eros$ similarity value as shown below,
\begin{equation}
\begin{aligned}
    D_{Eros}(A, B,W) &= \sqrt{2 -2Eros(A,B,w)} \\[1.2em]
    &= \sqrt{2 - 2\sum_{i = 1}^n w_i |<a_i, b_i>|} \\[1.2em]
    &= \sqrt{2 - 2\sum_{i = 1}^n w_i |\sum_{j=1}^n a_{ij} \times b_{ij}|}
\end{aligned}
\end{equation}

Essentially, we have a database of MTS samples where each MTS item represents the telemetry collected for a particular workload. There will be several such collections for each workload in the database. This database of MTS is first transformed into the eigenvector matrix space, that is each MTS item in the database is mapped to its eigenvector matrix and finally, in this space, we utilize the Eros distance metrics and modified leave-one-out KNN to evaluate the telemetry data and plot a precision vs recall graph.

Now that we have the database of different workloads with several collections of data when a new known workload’s telemetry is given we add that into the database and then transform it to the eigenvector matrix space, then we obtain the nearest neighbors to this workload and thus we will get to know that to what workloads is the unknown workload closer.

\section{Experiments and Results}
\label{sec:res}

\subsection{Data Collection}
For our experiments, we collected data for several different workloads in the NAS parallel benchmarks~\cite{NAS} and Intel Memory Latency Checker~\cite{MLC}, whereas the STREAM benchmark~\cite{stream} is kept aside and used as a test workload, that is, we see it as an unknown workload and then use the data collected over it to see against the already available workloads’ telemetry.
Table~\ref{tab:os_data} and Table~\ref{tab:hw_data} provide a sample of the collected data to understand the format of the data. This data is collected using the Intel Memory Latency Checker workload on a machine with Intel(R) Core(TM) i5-8250U CPU @ 1.60GHz over all 8 cores. Table~\ref{tab:os_data} shows a sample of the sar data collected for MLC benchmark. First row in the table represents the timestamp for each sample as 'Unix epoch'. Every other column in the table represents a metric reported by sar. We are showing a subset of total metrics collected by sar in Table~\ref{tab:os_data}. Each row in the table to shows the metrics values collected at a given time stamp when the workload was the exclusive application running on the system.  The data is sampled per second i.e., each row represents an interval of 1 second of workload execution. Only five rows are shown here for explanation purposes. For example, the third row in column ``\%user" represents the CPU utilization in user mode at timestamp ``1632834011".
 Similarly Table~\ref{tab:hw_data} shows a sample of the linux perf data collected for MLC benchmark. The timestamp in linux perf output are not in Unix epoch. We used addition post processing to convert the timetsamps to Unix epoch using collection start time. For example, the second row in column ``branches" represents the number of branches at timestamp ``2" seconds since the start of the collection.

\subsection{Experimental Results}
We collected the telemetry data using SYSSTAT (SAR) ~\cite{SAR} and 'linux perf' separately for each of the specified workloads 9 times. We run our technique on the collected data and measure the similarity of each pair of selected workloads.
Table \ref{tab:sim_sar} and Table \ref{tab:sim_perf} show the similarity results in the form of the Eros distance metric for every pair of workloads in a single collection. 
We can observe that the diagonal elements representing the distance between pair of same workloads are zero.


\begin{scriptsize}
\begin{table*}[h!]
    \centering
    \caption{sample distance matrix for a single collection of data using SYSSTAT (SAR) ~\cite{SAR}}
    \label{tab:sar_dist}
    \begin{tabular}{|l|l|l|l|l|l|l|l|l|l|}
    \hline
    &  $BT_B$ &  $BT_C$ &  $CG_C$ &  $FT_C$ &  $LU_B$ &  $LU_C$ &   MLC &  $SP_B$ &  $SP_C$ \\
    \hline
    $BT_B$ &  0.00 &  0.32 &  1.06 &  1.39 &  0.39 &  0.38 &  0.56 &  0.35 &  0.23 \\
    \hline
    $BT_C$ &  0.32 &  0.00 &  1.21 &  1.37 &  0.30 &  0.19 &  0.72 &  0.29 &  0.19 \\
    \hline
    $CG_C$ &  1.06 &  1.21 &  0.00 &  1.29 &  1.28 &  1.26 &  0.60 &  1.25 &  1.16 \\
    \hline
    $FT_C$ &  1.39 &  1.37 &  1.29 &  0.00 &  1.35 &  1.35 &  1.33 &  1.35 &  1.39 \\
    \hline
    $LU_B$ &  0.39 &  0.30 &  1.28 &  1.35 &  0.00 &  0.26 &  0.79 &  0.17 &  0.33 \\
    \hline
    $LU_C$ &  0.38 &  0.19 &  1.26 &  1.35 &  0.26 &  0.00 &  0.77 &  0.24 &  0.23 \\
    \hline
    MLC &  0.56 &  0.72 &  0.60 &  1.33 &  0.79 &  0.77 &  0.00 &  0.75 &  0.66 \\
    \hline
    $SP_B$ &  0.35 &  0.29 &  1.25 &  1.35 &  0.17 &  0.24 &  0.75 &  0.00 &  0.27 \\
    \hline
    $SP_C$ &  0.23 &  0.19 &  1.16 &  1.39 &  0.33 &  0.23 &  0.66 &  0.27 &  0.00 \\
    \hline
    \end{tabular}
\label{tab:sim_sar}
\end{table*}
\end{scriptsize}

\begin{scriptsize}
\begin{table*}[h!]
    \centering
    \caption{sample distance matrix for a single collection of data using Linux Perf~\cite{perf}}
    \label{tab:perf_dist}
    \begin{tabular}{|l|l|l|l|l|l|l|l|l|l|}
    \hline
    &  $BT_B$ &  $BT_C$ &  $CG_C$ &  $FT_C$ &  $LU_B$ &  $LU_C$ &   MLC &  $SP_B$ &  $SP_C$ \\
    $BT_B$ &  0.00 &  0.52 &  0.64 &  0.66 &  0.41 &  0.21 &  0.38 &  0.54 &  0.19 \\
    \hline
    $BT_C$ &  0.52 &  0.00 &  0.33 &  0.19 &  0.66 &  0.55 &  0.40 &  0.92 &  0.44 \\
    \hline
    $CG_C$ &  0.64 &  0.33 &  0.00 &  0.34 &  0.79 &  0.66 &  0.36 &  1.00 &  0.57 \\
    \hline
    $FT_C$ &  0.66 &  0.19 &  0.34 &  0.00 &  0.77 &  0.67 &  0.52 &  1.01 &  0.60 \\
    \hline
    $LU_B$ &  0.41 &  0.66 &  0.79 &  0.77 &  0.00 &  0.42 &  0.59 &  0.53 &  0.47 \\
    \hline
    $LU_C$ &  0.21 &  0.55 &  0.66 &  0.67 &  0.42 &  0.00 &  0.43 &  0.47 &  0.29 \\
    \hline
    MLC &  0.38 &  0.40 &  0.36 &  0.52 &  0.59 &  0.43 &  0.00 &  0.73 &  0.29 \\
    \hline
    $SP_B$ &  0.54 &  0.92 &  1.00 &  1.01 &  0.53 &  0.47 &  0.73 &  0.00 &  0.62 \\
    \hline
    $SP_C$ &  0.19 &  0.44 &  0.57 &  0.60 &  0.47 &  0.29 &  0.29 &  0.62 &  0.00 \\
    \hline
    \end{tabular}
\label{tab:sim_perf}
\end{table*}
\end{scriptsize}

\begin{scriptsize}
\begin{table}[h!]
  \centering
  \caption{details of the created SYSSTAT (SAR) ~\cite{SAR} database}
  \label{tab:sar_database}
  \begin{tabular}{|l|l|}
    \hline
    \textbf{Database Parameters} & \textbf{Value/Count} \\
    \hline
    \textbf{\# of variables} & 29\\
    \hline
    \textbf{the average length of the MTS} & 128\\
    \hline
    \textbf{\# of labels} & 9\\
    \hline
    \textbf{\#  of samples per label} & 9 \\
    \hline
    \textbf{total \# of samples} & 81\\
    \hline
  \end{tabular}
\end{table}
\end{scriptsize}

\begin{scriptsize}
\begin{table}[h!]
  \centering
  \caption{details of the created Linux Perf~\cite{perf} database}
  \label{tab:perf_database}
  \begin{tabular}{|l|l|}
    \hline
    \textbf{Database Parameters} & \textbf{Value/Count} \\
    \hline
    \textbf{\# of variables} & 8\\
    \hline
    \textbf{the average length of the MTS} & 125\\
    \hline
    \textbf{\# of labels} & 9\\
    \hline
    \textbf{\#  of samples per label} & 9 \\
    \hline
    \textbf{total \# of samples} & 81\\
    \hline
  \end{tabular}
\end{table}
\end{scriptsize}

\begin{figure}[ht]
\centering
\includegraphics[width=3.3in]{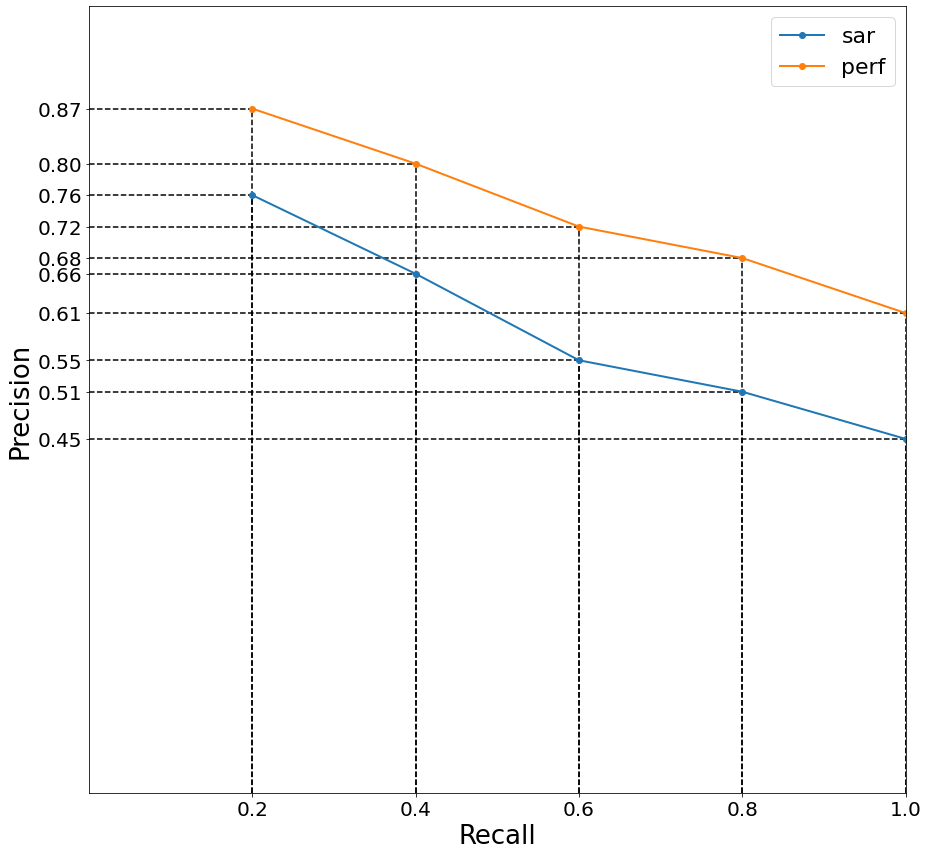}
\caption{Precision Vs Recall with modified KNN on the SYSSTAT (SAR) ~\cite{SAR}  and Linux Perf~\cite{perf}f database} 
\label{fig-pr}
\end{figure}

We follow~\cite{yang2004pca} and apply the modified leave-one-out KNN algorithm on the collected data (summarized in Table~\ref{tab:sar_database} and Table~\ref{tab:perf_database}) to obtain the precision vs recall curves, which is shown in Fig.~\ref{fig-pr}. 
From the set of MTS items in the collected data, we consider each workload as a query item and go through the whole database, leaving the query item, while using the modified KNN search to obtain the precision values. We use $r$ to represent the number of relevant items for the given query, which also acts like the $TP$ (true positive) value. Accordingly for each query and $r$ the corresponding recall values are obtained as $(=r/maxr)$ where $maxr$ stands for maximum relevant items. 
For every query item, we start with $r = 1$ and $k = 1$. We keep increasing $k$ (for modified KNN search) until we find exactly $c = r$ items that are of the same label as query item. At this point we get precision $(p) = c/k$ (for $c = r$) which we aggregate for all the query items for that particular value of $r$. 
Each of these aggregates are divided by $N$ (total number of MTS items in the database) to obtain the average precision value for each value of $r$ (or each value of recall value).

The precision vs recall curves in Fig. \ref{fig-pr} help in validating the effectiveness of the distance metric (Eros) on a given database of MTS items. The better the metric is able to distribute the data points (MTS items) in the search space, the better the modified KNN algorithm will be able to find the nearest neighbors that are of same label as the query item. The search space we have is modeled using the Eros distances between the MTS items.  We will have highest possible precision value $p = 1.0$ for $c = k = r = TP$, this is nothing but the case when $FP = 0$.

The precision value obtained signifies the relevance of the of the retrieved MTS items with respect to the given label. The combination of modified KNN and Eros together is used to retrieve the $maxr$(=1 to 5) number of items and the precision value explains how well it was able to retrieve the same label items on average. The downward trends in Fig. \ref{fig-pr} show the decay in the ability of the  modified KNN + Eros combination in finding the same label items as we try to retrieve more number of same label items for a given label. The PR curves serve as a way to compare the data distributions obtained using Sar and Perf.

\begin{figure}[ht]
\centering
\includegraphics[width=3.3in]{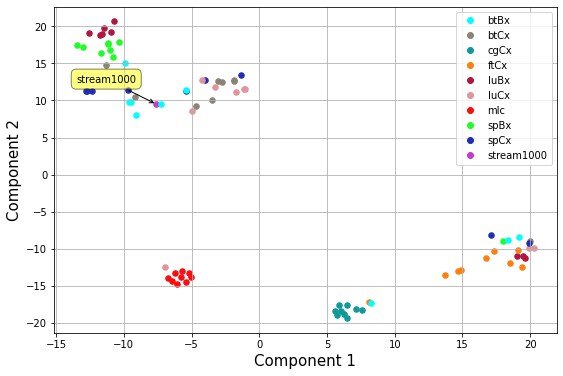}
\caption{t-SNE 2D Scatter Plot of SYSSTAT (SAR) ~\cite{SAR} Data using Eros as Distance metric} 
\label{fig-scatter-sar}
\end{figure}

\begin{figure}[ht]
\centering
\includegraphics[width=3.3in]{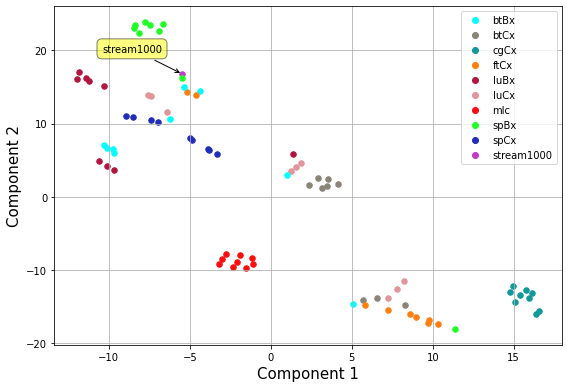}
\caption{t-SNE 2D Scatter Plot of Linux Perf~\cite{perf} Data using Eros as Distance metric} 
\label{fig-scatter-perf}
\end{figure}

Additionally, we take telemetry data of a new unknown workload that is not present in the database and try to find a know workload similar to the new unknown workload. In this case, we take the telemetry data of the STREAM benchmark and try to check it against the other workloads. We use t-SNE 2D scatter plot to build a visualization to see where the unknown workload falls. Fig. \ref{fig-scatter-sar} and Fig. \ref{fig-scatter-perf} are visualizations built using the t-SNE algorithm. We use the pair-wise Eros distances between the workload telemetry MTS data collections as precomputed metric and project the data space onto a two-dimensional plane. As stated previously, we have 9 collections for each workload and the newly taken STREAM workload is shown.

The difference in the PR curves is due to ability of KNN to distinguish between different workload telemetry data. The PR curve obtained on Perf data is relatively better compared to the one obtained on Sar because of the quality of the data obtained using Perf. In case of SYSSTAT (SAR) ~\cite{SAR}, in the visualizations, in Fig.~\ref{fig-scatter-sar} there is much more overlap or closeness between the data points as compared to the Fig.~\ref{fig-scatter-perf} 



\section{Conclusion}
The results obtained from the leave-one-out KNN on our database are for only nine collections and thus the results are moderately good. Doing this for many more collections might improve the performance of precision vs recall. This methodology is based on a relatively simple similarity metric and the way data is collected might have some flaws. For further work, we would want to explore deep metric learning instead of traditional machine learning and use deep learning methods for similarity analysis.
\bibliographystyle{IEEEtranS}
\bibliography{refs}

\end{document}